\documentclass[a4paper,12pt]{article}
\usepackage{amsmath}
\usepackage[dvips]{graphicx}
\newcommand {\Vect} [1] {{\bf #1}}
\newcommand {\Tens} [1] {{\bf #1}}
\newcommand {\eq} [1]   {(\ref{#1})}

\newcommand {\dfeq}     {\stackrel{\mbox{\scriptsize def}}{=}}
\newcommand {\be} [1]   {\begin{equation}\label{#1}}
\newcommand {\ee}       {\end{equation}}
\newcommand {\LL }  {\left(}
\newcommand {\RR }  {\right)}

\newcommand {\e}  { \Vect{e} }
\newcommand {\TU }  {\widetilde{U}}
\let\DS     = \displaystyle
\def\R{\Vect{R}}
\def\F{\Vect{F}}

\begin{document}
\bigskip
\begin{center}
{\LARGE
 Comment on the calculation of forces for multibody interatomic potentials}
 \\ [4mm]
\large
Vitaly A. Kuzkin
 \\ [4mm]
 Institute for Problems in Mechanical Engineering RAS \\ [4mm]
\end{center}
\bigskip
\bigskip
\begin{abstract}
The system of particles interacting via  multibody interatomic potential of general form is considered.
Possible variants of partition of the total force acting on a single particle into pair contributions are discussed.
Two definitions for the force acting between a pair of particles are compared.
The forces coincide only if the particles interact via pair or embedded-atom potentials.
However in literature both definitions are used in order to determine Cauchy stress tensor.  A simplest example of the linear pure shear of perfect square lattice is analyzed.
 It is shown that, Hardy's definition for the stress tensor gives different results depending on the radius of localization function. The differences strongly depend on the way of the force definition.

\end{abstract}

\section{Introduction}
Classical molecular dynamics~(MD)~\cite{Hoover_MD, Allen}, based on the
numerical solution of the Newtonian equations of motion of many interacting
particles, has been widely used for physical modeling and simulation for
several decades. Interactions between particles are usually described  by
so-called empirical interatomic potentials. A variety of potentials has been
proposed: starting with simple pair potentials~\cite{LJ1}, more accurate
embedded-atom potential~\cite{DawBaskes_1984} and ending with complex
bond-order potentials~\cite{SW, Tersof, Brenner(1990)}. The MD simulation
technique requires calculation of the total force acting on every particle.
If the potential energy is known then force~$\F_i$ acting on particle number~$i$ is determined by the following formula~\cite{LL}
\be{e1.2}
\F_i = -\frac{\partial U(\{\R_i\}_{i=1}^N)}{\partial \R_i} \ ,
\ee
 where $\{\R_i\}_{i=1}^N$ is the set of radius-vectors of particles; $N$ is the total number of particles; $U$ is the total potential energy of the system.
  Thus formula~\eq{e1.2} is sufficient for MD simulation. However the problem of interpretation and verification of  results  of the simulation is not so straightforward. The usual way is to calculate continuum variables, such as stress tensor and heat flux, during MD simulation and compare the results with the predictions of continuum theory. This problem was addressed by many authors starting from the late 19th century~\cite{Clausius}. Comprehensive reviews on this topic may be found in papers~\cite{Zhou2003, Zimmerman2004}. The majority of approaches use different energy and forces assumptions. Though the correctness of these assumptions is clear for pair potentials, for multibody potentials,
such as Tersoff potential, this is no longer the case. In particular, in Hardy's  formalism~\cite{Hardy} it was assumed that the total force can be
expressed by the summation of pair contributions~$\F_{ij}$ satisfying Newton's third law, i.e.
\be{e1.3}
\F_i  = \sum_{j\neq i} \F_{ij}, \qquad \F_{ij} = -\F_{ji}.
\ee
Here and below the summation is carried out over all particles in the system.
This assumption was examined in paper~\cite{JCP}. It was  shown that it holds
for systems with three-body forces~\cite{SW, Tersof, Brenner(1990)}. The
generalization for N-body forces was carried out in~\cite{Delph}.  In
paper~\cite{Zimmerman2004} it was stated that the partition~\eq{e1.3} can
always be carried out. However, the physical meaning of~$\F_{ij}$ is clear
only in the case of pair potentials. One can easily understand the reason for
this ambiguity. Systems with pair interactions are simply equivalent to sets
of particles connected by longitudinal springs. Therefore~$\F_{ij}$ is just a
force caused by the deformation of the spring. In contrast the simplest
systems with multibody interactions can be imagined as a set of particles
connected by longitudinal and angular springs between the bonds. In this case every angular spring belongs
to three particles and causes the forces acting on all of them. Even in this
simple case partition~\eq{e1.3} is not straightforward. Note that commonly
used multibody potentials, like Tersoff~\cite{Tersof}, are even more complex.
Another definition for~$\F_{ij}$ was given in papers~\cite{Kuzkin} and
\cite{Zimmerman2010}.  It was stated in~\cite{Zimmerman2010} that division~\eq{e1.3} may be non-unique. However this statement was not proved.

In the present paper the problem of the partition of the total force into pair
contributions is discussed in detail. Two definitions for the force~$\F_{ij}$
acting between a pair of particles~(material points)\footnote{Note that particles with rotational degrees of freedom, internal structure, etc. are not considered.} are considered. It is shown that the partition is not
unique and does depend on the way the potential energy is represented.  The influence of the partition on the value of stresses is analyzed  for the simplest example --- a linear pure shear of a square lattice.

\section{Comparison of different definitions for the force acting between two particles}
Let us begin with consideration of the results obtained in papers~\cite{JCP, Delph}.
The problem of calculation of~$\F_{ij}$ in the case of three-body forces was
addressed in~\cite{JCP}. It was shown that  both  the Stillinger-Weber~\cite{SW} and the
Tersoff~\cite{Tersof} potentials can be expressed
in the following form
\be{e4}
 U = \frac{1}{2}\sum_{i} U_i, \quad U_i = \sum_{j\neq i;k \neq i,j} U_{ijk}(R_{ij}, R_{ik},
 R_{kj}), \quad R_{ij} = |\R_{ij}|.
\ee
Here and below $\R_{ij} = \R_i-\R_j$. Then according to  definition~\eq{e1.2} the total force acting on
the $i$-th particle is
\be{e7}
  \Vect{F}_i  = \sum_{j \neq i} \F_{ij}, \qquad \F_{ij} \dfeq  -\frac{1}{2} \sum_{k \neq i,j} \frac{\partial (U_{ijk}+U_{kij} + U_{kji} + U_{jik}+U_{ikj} + U_{jki})}{\partial R_{ij}}\e_{ij}.
\ee
Here and below~$\e_{ij} \dfeq  \R_{ij}/R_{ij}$.
The definition for force~$\F_{ij}$ acting between particles $i$ and $j$ arises in a
natural way while calculating the derivative. On the other hand in paper~\cite{JCP} it was shown
 that the same results can be obtained using the following definition
\be{e8}
    \F_{ij} \dfeq  -\frac{\partial U}{\partial R_{ij}} \e_{ij}.
\ee
One can see that equations~\eq{e1.3} are satisfied and furthermore~$\F_{ij}$ are central, i.e. parallel to the vector connecting particles~$i$ and $j$. The above mentioned approach was generalized in paper~\cite{Delph}. Let us assume that the total potential energy depends on all
the interatomic distances in the system, i.e.
\be{e8.1}
 U = U(\{R_{kn}\}_{k,n>k}).
\ee
Formula \eq{e8.1} is the most general form for potential energy of the atomic system.
According to definition~\eq{e1.2} the total force~$\F_i$ is
\be{}
  \F_{i} = -\sum_{k,j>k} \frac{\partial U}{\partial R_{kj}}\frac{\partial R_{kj}}{\partial \R_{i}} =  -\sum_{j \neq i} \frac{\partial U}{\partial R_{ij}}  \e_{ij}.
 \ee
One can see that again the definition~\eq{e8} naturally follows from the derivation. Thus in the general case~\eq{e8.1} the total force can be expressed in form~\eq{e1.3}.


 The partition discussed above requires the potential energy~$U$ to be
 represented as a function of all the interatomic distances in the
 system~\eq{e8.1}. The requirement is automatically satisfied in the case of
 pair potentials~\cite{LJ1} and the embedded-atom potential~\cite{DawBaskes_1984}.
 However representation~\eq{e8.1} may be inconvenient in the case of bond-order potentials, such as Stillinger-Weber~\cite{SW}
 and Tersoff~\cite{Tersof}, which depend on the angles between bonds.
 Let us consider the different approach proposed in papers~\cite{Kuzkin} and \cite{Zimmerman2010}.
 Assume that the total potential energy of the system has the form
\be{e1}
  U = \sum_i U_i(\{\R_{ij}\}_{j \neq i}).
\ee
Obviously all potentials mentioned above can be represented in the form~\eq{e1}.
In contrast to the previous approach, the geometry of the atomic system should be represented via vectors~$\{\R_{ij}\}_{j \neq i}$,
but not interatomic distances. At first glance it seems that both approaches are equivalent. Let us show that in general it is not true.
Calculating the total force~$\F_i$
using the definition~\eq{e1.2} and expression~\eq{e1} one obtains
\be{1}
 \DS  \F_i =  -\sum_j  \sum_{k \neq j} \frac{\partial U_j}{\partial  \R_{jk}}\cdot \frac{\partial \R_{jk}}{\partial \R_{i}} = \sum_{j\neq i} \LL   \frac{\partial U_j}{\partial\R_{ji}} - \frac{\partial
  U_i}{\partial\R_{ij}}\RR, \quad \Vect{F}_{ij} \dfeq \frac{\partial U_j}{\partial\R_{ji}} - \frac{\partial U_i}{\partial\R_{ij}}.
 \ee
One can see that introduced definition for~$\Vect{F}_{ij}$ satisfies Newton's
 third law, i.e.~$\F_{ij}=-\F_{ji}$.
Therefore~$\Vect{F}_{ij}$ may be considered as a force acting between
particles $i$ and $j$. It can be shown that, in the case of pair potentials and embedded-atom potentials, forces~$\Vect{F}_{ij}$ are central.  Moreover in these particular cases definitions~\eq{e8}  and~\eq{1} exactly coincide. However in general this is not true.  The first distinction and, probably, the most important one is that the partition of the total energy~\eq{e1}, used in the expression~\eq{1}, is not unique. According to formula~\eq{1} the partition can, in general, affect the value of the force~$\F_{ij}$. However the majority of commonly used potentials are based on partition~\eq{e1} as well. For example, in the case of Tersoff and Stillinger-Weber potentials the partition is determined by formula~\eq{e4}. The comparison of different partitions will be carried out below for a simple example.


Let us compare definitions~\eq{e8}, \eq{1} in the case of three-body potentials.
Rewriting the expression for potential energy~\eq{e4} in the form analogous to~\eq{e1} one obtains
\be{e9}
  U = \sum_i U_i, \quad U_i = \frac{1}{2}\sum_{k, n \neq k} \TU_{ikn}(\R_{ik}, \R_{in}), \quad \TU_{ikn}(\R_{ik}, \R_{in}) = U_{ikn}(R_{ik}, R_{in}, R_{kn}).
  \ee
One can show that in this case the expression~\eq{1} for $\F_{ij}$ takes the form
\be{e10}
 \DS \Vect{F}_{ij} \dfeq \frac{1}{2}\sum_{k \neq i,j} \left[ \frac{\partial }{\partial\R_{ji}}\LL \TU_{jik} + \TU_{jki} \RR -  \frac{\partial }{\partial\R_{ij}}\LL \TU_{ijk} + \TU_{ikj} \RR\right].
 \ee
Let us substitute the last  formula from~\eq{e9} into~\eq{e10} and calculate the derivatives.
 Then~$\F_{ij}$ can be expressed in the form analogous to~\eq{e7}
\be{e13}
  \Vect{F}_{ij} \dfeq -\frac{1}{2}\sum_{k \neq i,j} \left[ \frac{ \partial \LL U_{ijk} + U_{ikj}  + U_{jik}  + U_{jki}\RR}{\partial  R_{ij}}  \e_{ij}  + \frac{\partial \LL U_{jik} + U_{jki}\RR}{\partial R_{ik}} \e_{ik}  +    \frac{\partial \LL U_{ijk} + U_{ikj}\RR}{\partial R_{kj}}  \e_{kj} \right].
\ee
One can clearly see that expressions~\eq{e7} and~\eq{e13}  are different.
In particular, from~\eq{e7} it follows that forces $\F_{ij}$ are parallel
to~$\R_{ij}$. According to~\eq{e13} this is not true.  Further these
statements will be explicitly shown for a simple example~(see equations~\eq{e20}, \eq{e17}, \eq{e17_Zim}).
\section{Calculation of Cauchy stress tensor. The simplest example}
 Let us consider the practical consequences of different ways of partitioning of the total force. It was mentioned above that results of MD simulation are usually compared with predictions of continuum theory. Equivalent continuum quantities, such as Cauchy stress tensor, are calculated for this purpose. The stress tensor characterizes external forces acting on the material surface surrounding the volume selected from the continuum media. Therefore the forces should be divided into internal and external in order to calculate stress. The analogous situation occurs if one defines equivalent stress tensor for discrete system. Obviously it is impossible to express stress tensor in terms of the total forces acting on the particles. Therefore partition~\eq{e1.3} is required. It was shown above that the partition is not unique. Thus we obtain the result that the stresses in a material can depend on the choice of definition for~$\F_{ij}$. Let us consider the simplest example, notably linear pure shear of a perfect zero-temperature  square lattice. Let particles interact via angular springs connecting two neighboring bonds. Only nearest neighbors will be taken into account. Longitudinal springs may also be added but their contribution to forces~$\F_{ij}$ is the same for both expressions~\eq{e8}, \eq{1}. Let us consider particle number~0 with radius-vector~$\R_0$ and
denote its neighbors as in figure~1.
\begin{figure}[!ht]
\centering
\begin{minipage}[t]{0.35\textwidth}
\includegraphics[width=\textwidth]{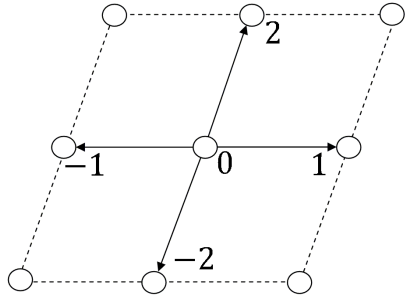}
\caption{Particle number~$0$ and its nearest neighbors.}
 \label{fig-1}
\end{minipage}
\end{figure}
Note that this numbering takes symmetry into account, for example, $\R_{01} = -\R_{0(-1)}$.
Let us assume that potential energy of the spring connecting bonds~$\R_{01}$ and~$\R_{02}$ is given by the following formula
\be{ne1}
 U_{012} = \frac{c}{2}\LL \e_{01}\cdot\e_{02} \RR^2 = \frac{c}{2} \LL\frac{R_{01}^2 +  R_{02}^2 - R_{12}^2}{2 R_{01} R_{02}} \RR^2.
\ee
 In the case of small deformations formula~\eq{ne1} corresponds to the energy of harmonic angular spring with stiffness~$c$.
Let us derive expressions for forces~$\F_{01}$, $\F_{02}$, $\F_{12}$, $\F_{1(-2)}$. The remaining forces can be obtained using
 symmetry or Newton's third law, for example, $\F_{12}=\F_{(-2)(-1)}, \F_{01}=-\F_{10}$.
First let us use definition~\eq{e8}. Substituting formula~\eq{ne1} into formula~\eq{e7} and linearizing  the resulting expression in the case of small
shear~$(\varphi \dfeq {\rm arcsin} (\e_{01} \cdot \e_{02}), \varphi \ll 1)$ one obtains
\be{e20}
\F_{01} \approx 0, \quad  \F_{02} \approx 0, \quad \F_{12} \approx \frac{2\sqrt{2}c \varphi}{R}\e_{12},
\quad \F_{1(-2)} \approx \frac{2\sqrt{2}c \varphi}{R}\e_{1(-2)}.
\ee
where $R = R_{01} = R_{02}$. Here and below terms of order higher then~$\varphi$ are neglected. It is clear that forces~\eq{e20} are central.

Let us use definition~\eq{1}. It was mentioned above that representation~\eq{e1} of the total energy used in formula~\eq{1} is not unique. In the case under consideration particle number~0 is surrounded by four nearest neighbors and four corresponding spring.
Let us compare two possible ways of the partition of the total energy. First, let us assume that the energy of the springs contribute to energy~$U_0$ of particle number~0 only. This way of partition is the simplest one and it is similar to the partition used in the definition for Tersoff and Stillinger-Weber potentials~\eq{e9}. Calculating forces~$\F_{01}$, $\F_{02}$, $\F_{12}$, $\F_{1(-2)}$ in the framework of the assumptions used during the derivation
of formula~\eq{e20} one obtains
\be{e17}
 \F_{01} \approx -\frac{4c\varphi}{R} \e_{02}, \quad \F_{02} \approx -\frac{4c\varphi}{R} \e_{01}, \quad \F_{12} = \F_{1(-2)} = 0.
\ee
Obviously  the forces~\eq{e17} are non-central and differ from the previous
 result~\eq{e20}. Note that in formula~\eq{e17} forces~$\F_{12}, \F_{1(-2)}$ are exactly equal to zero. The second way of partition
 of the total energy was proposed in paper~\cite{Zimmerman2010}. Energy~$U_{012}$ of the spring was divided between particles
  number 0, 1 and 2 in equal portions. Calculating the forces one obtains
\be{e17_Zim}
 \F_{01} \approx -\frac{8c\varphi}{3R} \e_{02}, \quad \F_{02} \approx -\frac{8c \varphi}{3R} \e_{01}, \quad \F_{12} \approx \frac{2\sqrt{2}c \varphi}{3R}\e_{12},
\quad \F_{1(-2)} \approx \frac{2\sqrt{2}c \varphi}{3R}\e_{1(-2)}.
\ee
Note that in contrast to formula~\eq{e17}, forces~$\F_{12}, \F_{1(-2)}$ determined by formula~\eq{e17_Zim} are not equal to zero.


 Let us calculate shear stress acting on cross-section with normal~$\e_{01}$ of the crystal.  According to classical continuum mechanics definition, the shear stress acting on the cross-section is equal to the component of the force parallel to vector~$\e_{02}$ per unit length. One can prove using formulas~\eq{e20}, \eq{e17}, \eq{e17_Zim} that in all the cases
the absolute value of shear stress~$\tau$ can be expressed as
\be{shear stress}
 \tau = C_{44}^* |\varphi|, \quad C_{44}^* \dfeq \frac{4c}{R^2}.
\ee
The second formula from~\eq{shear stress} coincides with the well-known expression for elastic constant~$C_{44}$ of the square lattice, obtained, for example, in~\cite{Krivtsov}. Index star is used in order to mark the exact solution. Thus, at least in the case under consideration, stresses calculated using different expressions for~$\F_{ij}$ coincide.

Application of mentioned above definition for the stress is not very convenient, especially for dynamical problems involving large thermal motion, structural transformations, fracture, etc. Therefore in practice different approaches for calculation of stress are used~(see papers \cite{Zhou2003, Zimmerman2004} for detailed reviews). In particular, Hardy's formalism~\cite{Hardy} is frequently used~\cite{Zimmerman2004, JCP, Zimmerman2010}. In papers~\cite{JCP, Zimmerman2010} it was shown that in the framework of Hardy's
formalism the potential part of Cauchy stress tensor at spatial point~$\R_0$ has the form
\be{e14}
 \Tens{\sigma}_{pot}(\R_0, t) = -\frac{1}{2} \sum_i \sum_{j\neq i} \F_{ij} \R_{ij} B_{ij}(\R_0),  \qquad B_{ij}(\R_0) \dfeq \int_0^1 \psi(\lambda \R_{ij} + \R_{j} - \R_0) d \lambda,
\ee
where~$\psi$ is so-called localization function~(see~\cite{Zimmerman2010} for details).   Formula~\eq{e14} was used in paper~\cite{JCP} and in the Appendix of paper~\cite{Zimmerman2010}. However the quantity~$\F_{ij}$ has
different meanings in them. Formula~\eq{e8} was used as a definition for~$\F_{ij}$ in paper~\cite{JCP}. In contrast in paper~\cite{Zimmerman2010}
 quantity~$\F_{ij}$ was calculated  using formula~\eq{1}. It was shown above that, in general, forces calculated with the use of~\eq{e8} and \eq{1}
  are different. Thus corresponding stress tensors are also different.  The accurate comparison of the stress tensors  will be addressed in a separate paper. In the present work only the simplest example is analyzed.

Let us return to the example of linear pure shear of the square lattice.
Consider Cauchy stress tensor at the point~$\R_0$, where particle number~$0$ is placed. One can prove  substituting formulas~\eq{e20},  \eq{e17}, \eq{e17_Zim} into formula~\eq{e14} that in all these cases stress tensor is symmetric. Note that, in general, this is not
obvious as forces~\eq{e17} and~\eq{e17_Zim} are non-central.
 Let us calculate elastic constant~$C_{44}$ of the system as it completely determines stresses in the case under consideration.
 The exact solution of this problem is given by the second formula from~\eq{shear stress}.
For simplicity the radial step function was used as a localization function,
 i.e.~$\psi=1/(\pi r_L^2)$ inside the localization volume and zero in the
 remaining space, where~$r_L$ is  radius of the localization volume.
 The elastic constants calculated with different values of~$r_L$ are shown in figure 2.
\begin{figure}[!ht]
\centering
\begin{minipage}[t]{0.8\textwidth}
\includegraphics[width=\textwidth]{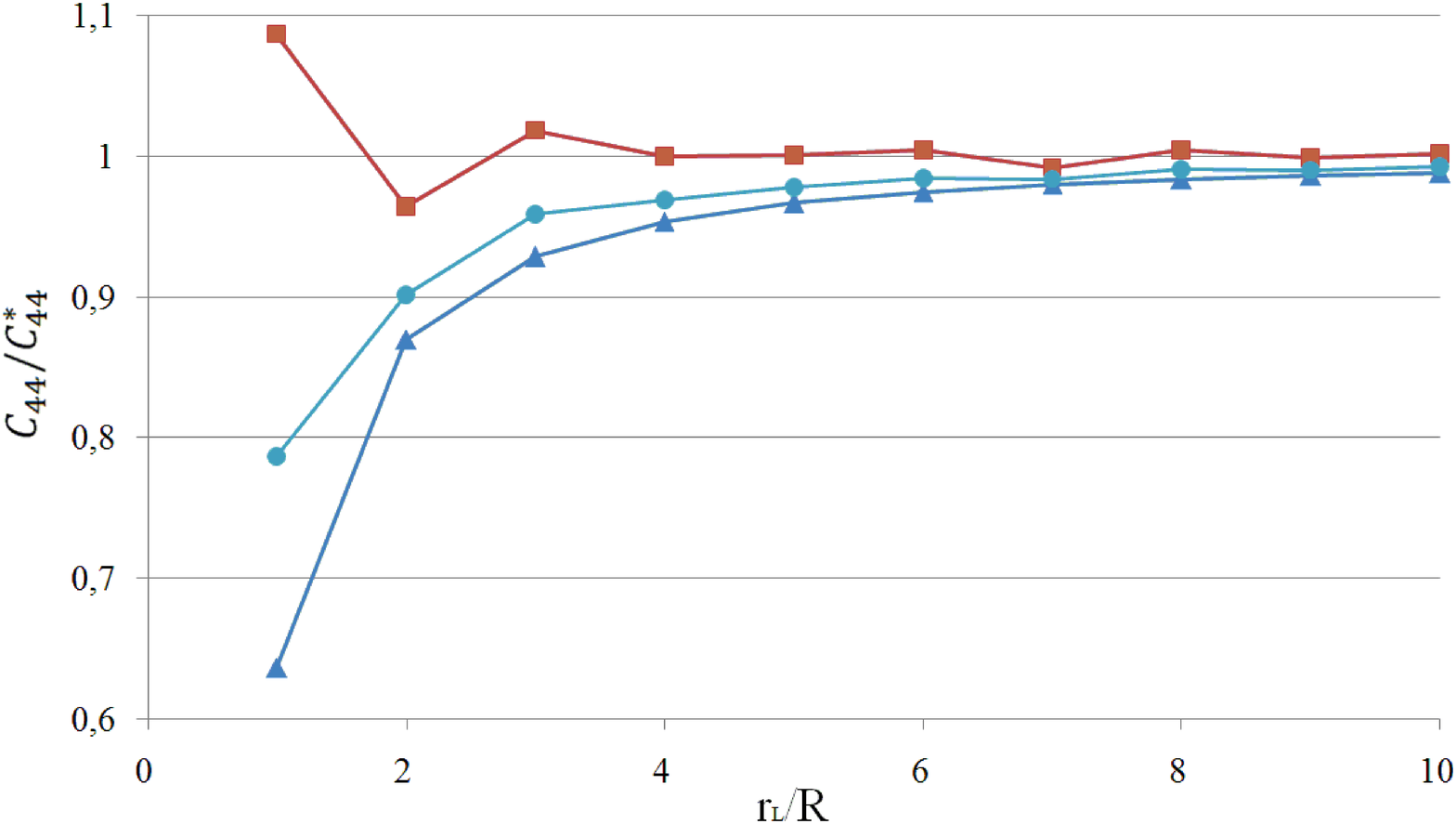}
\caption{Calculated elastic constant~$C_{44}$ divided by the exact solution~$C_{44}^*$.
Squares, triangles and circles correspond to elastic constant calculated using expressions~\eq{e20}, \eq{e17}, \eq{e17_Zim} respectively.}
 \label{fig-2}
\end{minipage}
\end{figure}
One can see from figure 2 that for small~$r_L/R$ elastic constants calculated using expressions~\eq{e8} and~\eq{1} for~$\F_{ij}$ are different. Furthermore in the second case elastic constant~$C_{44}$ depend on the partition of the total energy.
 However the expressions converge to the same value~$C_{44}^*$ with increasing radius of localization function. Note that the elastic constant which corresponds to definition~\eq{e8}
converges to the exact solution more rapidly then the elastic constant which corresponds to definition~\eq{1}. The practical consequence
of this fact is that the first approach requires a smaller value of~$r_L$ than
the second one and it is therefore more efficient from the computational point of
view.

\section{Results and discussion}
Let us summarize the results. The problem of the partition of forces acting in the discrete system with multibody interactions into pair contributions is analyzed.
Two methods for partitioning, differing in the way representation of potential energy~$U$, are discussed.
It is shown that in the framework of both methods the total force~$\F_{i}$ is expressed as a sum of pair contributions~$\F_{ij}$.
In both cases the definition for~$\F_{ij}$ is rather natural as the analog of
Newton's third law is satisfied, i.e. $\F_{ij} = -\F_{ji}$.
However the definitions  coincide only in the case of pair
potentials and the embedded-atom potential.
In particular, according to the approach proposed in papers~\cite{JCP, Delph}, forces $\F_{ij}$ are always central.
In the framework of the approach proposed in papers~\cite{Kuzkin, Zimmerman2010} forces~$\F_{ij}$ are, in general, non-central.
The difference was explicitly shown in the case of three-body potentials.
Thus the partition mentioned above is not unique. It depends on the
representation of the potential energy. The influence of the partition on the
value of stress tensor is analyzed for the simplest example of linear pure
shear of perfect zero-temperature square lattice.  Two definitions for stress are considered. If stresses are calculated as a force acting on the cross-section of the crystal per unit length, all expressions for~$\F_{ij}$ considered above
 lead to exactly the same value of stresses. In contrast, Hardy's definition~\eq{e14} gives different values of stresses depending on the size of localization volume and the differences depend strongly on the way of force definition.
Accurate comparison of different expressions for stress tensor in the case on nonlinear
deformation involving thermal motion is addressed in our future work.

Finally let us note that in the case of multibody interactions  a common
point of view on the definition for the force acting between two particles does not exist.
Summarizing the results of the present paper one can formulate several advantages of definition~\eq{e8}. First, definition~\eq{e8} does not explicitly depend on   partition~\eq{e1} of the total energy, which is, in general, not unique.
Secondly, it was mentioned that forces defined by formula~\eq{e8} are central. In this case equation of moment
of momentum balance is satisfied for any subsystem of the discrete system~\cite{Truesdell}. It leads, in particular, to unconditional symmetry of corresponding Cauchy stress tensor. Note that in the framework of classical continuum mechanics the symmetry is required~\cite{Truesdell}. On the other hand, in paper~\cite{Zimmerman2010} it was shown that definition~\eq{1} is more appropriate for the formulation of equivalent micromorphic continuum theory for discrete systems. The third appealing
feature of definition~\eq{e8} is that, at least in the considered case, the elastic constant~$C_{44}$ and therefore the stress tensor, calculated with the use of formula~\eq{e8}, converges to the exact solution more rapidly than in the case of definition~\eq{1}. In the last case it is shown that the convergence rate depend on the way of partition of the total energy between particles.

The author is deeply grateful to Prof. Y. Chen, Prof. Wm.G. Hoover, Prof. E.A. Ivanova, Prof. A.M. Krivtsov and Prof. J.A. Zimmerman
for useful discussions and inspiration.

\end{document}